\documentclass[a4paper,10pt]{article}
\usepackage[utf8x]{inputenc}

\title{A Superspace Formulation of the BV 
Action for Higher Derivative Gravity}
\author{
Mozzam Khan\\
Department of Mathematics,
King's College London\\
Strand,
London WC2R 2L}
\begin{document}

\maketitle

\begin{abstract}
In this paper we analyse perturbative higher derivative gravity  which is known to possess  a BRST symmetry
associated with its higher derivative
structure.  We first analyse the anti-BRST and double BRST symmetries of this theory.
 We then discuss the extended BRST and extended anti-BRST symmetries of this theory using the superspace formalism.
 We  show that even though this theory is generally invariant under extended BRST transformations under extended anti-BRST
transformations it is only invariant on-shell.
\end{abstract}



\section{Introduction}
It is well known that gravity is non-renormalisable \cite{nbg1}.   It is also known that 
higher derivative  theories have as a field theory better
renormalisation properties than conventional ones and, in particular, that the addition of higher derivative terms to the Lagrangian density for general relativity
 makes it renormalisable \cite{nbg2}. This motivates the study of gravity with higher derivative terms.

The validity of  general relativity  on cosmological scales has never been tested
\cite{ng1}. It is hoped that a generalization of  General Relativity theory may explain the accelerating universe.
\cite{ng2}-\cite{ng3}. This is the motivation for  the studying an $f(R)$-gravity which contains  higher derivatives.
These theories would also modify the gravitational potential.
It is hoped that a corrected gravitational potential could
fit galaxy rotation curves without the need of dark matter \cite{ng4}-\cite{ng5}. It is possible to work out a formal
analogy between the corrections to the Newtonian gravitational potential by the $f(R)$-gravity   and dark matter models.
The corrections to the Newton potential can be used to test the accuracy of these theories \cite{ng6}-\cite{ng7}.

However, the addition of higher derivative terms leads to the existence of negative norm states and this in turn breaks the
unitarity of these theories \cite{3a}-\cite{4a}.  Renormalisable models of gravity can be constructed by adding higher order spatial
 terms without higher order temporal terms, as is done in   Horava-Lifshitz gravity \cite{ng8}-\cite{ng9}.
However, this breaks the Lorentz invariance
of the resultant theory \cite{ng10}.

Negative norm states associated with the gauge symmetry are also known to occur as  Faddeev-Popov ghosts  in
Yang-Mills theories and it is possible to deal with them   by  means of BRST and anti-BRST symmetries  \cite{7a}-\cite{8a}.
BRST and anti-BRST symmetries of quantum gravity have also been analysed \cite{graz}-\cite{graz1}.
In fact,  gauge invariance 
 also has  very interesting consequences  in string theory \cite{za}-\cite{zb}. So, the BRST symmetry has been studied in string theory 
\cite{za1}-\cite{zb1}. 
It seems possible to use similar techniques to deal with ghosts associated with higher derivatives. In fact BRST 
symmetry associated with the higher derivative structure of   higher derivative  gravity theories has been recently studied~\cite{9a}-\cite{faizalmo}.

 In this paper we shall generalise the results of Ref. \cite{9a} to include anti-BRST symmetry and  double BRST symmetry.
Then we shall study the effect of shift symmetry on  perturbative higher derivative gravity in the
Batalin Vilkovisky (BV) formalism \cite{10a}-\cite{11a}.

The extended BRST and the extended anti-BRST symmetries of Faddeev-Popov ghosts \cite{12a}-\cite{13a} along
 with their superspace formalism are well understood  for conventional Yang-Mills theories \cite{14a}-\cite{15a}.
We shall apply these results to the ghosts associated with the higher derivative structure of the higher derivative gravity.

\section{BRST and Anti-BRST Invariant Lagrangian Density}
In this paper we shall only study perturbative gravity against a flat background metric.
We shall thus  split the full metric $g_{\mu\nu}^{(f)}$
into the metric for the background flat spacetime $\eta_{\mu\nu}$
 and a small perturbation around it, being $g_{\mu\nu}$,
\begin{equation}
 g_{\mu\nu}^{(f)} = \eta_{\mu\nu} + g_{\mu\nu}.
\end{equation}
The lowering and raising of indices are compatible with the metric for this background spacetime.
Now the Lagrangian density for  the higher derivative gravity theories with the
Lagrangian density $\mathcal{L}$ is given by
\begin{equation}
\mathcal{L}=  \frac{1}{2} \mathcal{O} g^{\mu\nu} \mathcal{O} g  _{\mu\nu} + \overline{c}^{\mu\nu}\mathcal{O} c_{\mu\nu}, \label{2a}
\end{equation}
where $c_{\mu\nu}$ is a ghost field, $\overline{c}_{\mu\nu}$  is an anti-ghost field, and $\mathcal{O}$
depends on the order of the theory.
Now, using an auxiliary field $L_{\mu\nu}$, we can   write the Lagrangian density  given in Eq. $(\ref{2a})$ as
\begin{equation}
\mathcal{L}=L^{\mu\nu}\mathcal{O} g    _{\mu \nu}-\frac{1}{2}L^{\mu\nu} L_{\mu\nu} + \overline{c}^{\mu\nu}\mathcal{O} c_{\mu\nu}. \label{4a}
\end{equation}
In Ref. \cite{9a} the  explicit example of a fourth order theory was studied.
For a  fourth order theory in flat spacetime  we have
\begin{equation}
L^{\mu\nu}\mathcal{O} g_{\mu \nu}= L^{\mu\nu}(a_1 R_{\mu\nu} + a_2 R g_{\mu\nu} + a_3 g_{\mu\nu} + S_{\mu\nu}), \label{3a}
\end{equation}
where $a_1, a_2, a_3$,  are constants, $R_{\mu\nu}$ is the Ricci curvature tensor,
 $R$ is the Ricci scalar obtained from $g_{\mu\nu}$ and $S_{\mu\nu}$ denotes the contribution coming from the gauge fixing terms. After
eliminating the auxiliary field this Lagrangian density  can be explicitly written as
\begin{eqnarray}
\mathcal{L}= \mathcal{L}_g + \mathcal{L}_{gf} + \mathcal{L}_{gh},
\end{eqnarray}
where $\mathcal{L}_g$ is the gravitational part of the Lagrangian density, $\mathcal{L}_{gf}$ is the contribution coming from the
gauge fixing part and $\mathcal{L}_{gh}$ is the contribution coming from the ghost part.
The gravitational part of the Lagrangian density  $\mathcal{L}_g$ will contain terms proportional to $R^2$
and $R^{\mu\nu}R_{\mu\nu}$~\cite{9a}.

We will refrain from defining the exact form of $\mathcal{O}$ in this paper so as to keep the results general
 and applicable to any  order gravity theory.
The  Lagrangian density  given by Eq. $(\ref{4a})$ is known to possess a BRST symmetry \cite{9a},
as it is  invariant under the following BRST transformations:
\begin{eqnarray}
\delta  g  _{\mu\nu}&=&c_{\mu\nu},\nonumber\\
\delta \overline{c}_{\mu\nu}&=&-L_{\mu\nu},\nonumber\\
\delta c_{\mu\nu}&=&0,\nonumber\\
\delta L_{\mu\nu}&=&0.
\end{eqnarray}
We note that the  Lagrangian density given by Eq $(\ref{4a})$ is also invariant under the following anti-BRST transformations:
\begin{eqnarray}
\overline{\delta} g  _{\mu\nu}&=&\overline{c}_{\mu\nu},\nonumber \\
\overline{\delta c}_{\mu\nu}&=&0,\nonumber \\
\overline{\delta}c_{\mu\nu}&=&L_{\mu\nu}, \nonumber \\
\overline{\delta}L_{\mu\nu}&=&0,\label{6a}
\end{eqnarray}
and so it can be written as
\begin{eqnarray}
\nonumber
\mathcal{L}&=&\overline{\delta} \left( c^{\mu\nu}\left(\mathcal{O} g  _{\mu\nu} -\frac{1}{2}L_{\mu\nu} \right) \right) \nonumber \\
&=&-\delta \left(\overline{c}^{\mu\nu}\left(\mathcal{O} g  _{\mu\nu}-\frac{1}{2}L_{\mu\nu} \right) \right) \nonumber \\
&=&\frac{1}{2}\overline{\delta}\delta  \left( g  ^{\mu\nu} \mathcal{O} g  _{\mu\nu}-c^{\mu\nu}\overline{c}_{\mu\nu} \right) \nonumber \\
&=&-\frac{1}{2}\delta\overline{\delta}  \left( g  ^{\mu\nu} \mathcal{O} g  _{\mu\nu} - c^{\mu\nu}\overline{c}_{\mu\nu} \right). \label{7aab}
\end{eqnarray}
Thus this Lagrangian density can be expressed not only as a total BRST or a total anti-BRST variation but also as a
 total double  BRST variation.
\section{Extended BRST Lagrangian Density}
We next consider the case of the extended BRST invariant Lagrangian density by first shifting the original fields as
\begin{eqnarray}
 g  _{\mu\nu} &\to&  g  _{\mu\nu} - \tilde{ g  }_{\mu\nu}, \nonumber \\
c_{\mu\nu} &\to& c_{\mu\nu} - \tilde{c}_{\mu\nu}, \nonumber \\
\overline{c}_{\mu\nu} &\to& \overline{c}_{\mu\nu} - \tilde{\overline{c}}_{\mu\nu},\nonumber\\
L_{\mu\nu} &\to&  L_{\mu\nu} - \tilde{ L}_{\mu\nu}.
\end{eqnarray}
The extended BRST invariant Lagrangian density is obtained by considering both the invariance of the original BRST transformations and the new shift
transformations of the original fields
\begin{equation}
\tilde{\mathcal{ L}}=\mathcal{L}( g  _{\mu\nu} - \tilde{ g  }_{\mu\nu}, c_{\mu\nu}- \tilde c_{\mu\nu}, \overline{c}_{\mu\nu}-\tilde{\overline c}_{\mu\nu}, L_{\mu\nu}- \tilde L_{\mu\nu}),
\label{9a}
\end{equation}
This extended BRST invariant Lagrangian density is given by Eq. $(\ref{9a})$ and is invariant under the following  extended BRST symmetry with the transformations
\begin{eqnarray}
\delta g  _{\mu\nu}=\psi_{\mu\nu}, && \delta \tilde g  _{\mu\nu}=(\psi_{\mu\nu}- (c_{\mu\nu}- \tilde c_{\mu\nu})),\nonumber\\
\delta c_{\mu\nu} =\epsilon_{\mu\nu}, && \delta \tilde{c}_{\mu\nu}=\epsilon_{\mu\nu},\nonumber \\
\delta \overline c_{\mu\nu} = \overline{\epsilon}_{\mu\nu},&&  \delta \tilde{\overline c}_{\mu\nu}=(\overline \epsilon_{\mu\nu} + (L_{\mu\nu}-\tilde L_{\mu\nu})), \nonumber\\
\delta L_{\mu\nu} =\rho_{\mu\nu}, && \delta\overline L_{\mu\nu}=\rho_{\mu\nu}.
\end{eqnarray}
Here, $\psi_{\mu\nu}, \epsilon_{\mu\nu}, \overline \epsilon_{\mu\nu} $ and $\rho_{\mu\nu}$ are the ghost fields associated with the shift symmetries of the original fields
 $ g  _{\mu\nu}, c_{\mu\nu},
 \overline {c}_{\mu\nu}$ and $L_{\mu\nu}$ respectively.
The  BRST transformations of these ghosts associated with the shift symmetry now vanish:
\begin{eqnarray}
\delta\psi_{\mu\nu} &=&0,\nonumber\\
\delta\epsilon_{\mu\nu} &=&0,\nonumber \\
\delta \tilde{\epsilon}_{\mu\nu}&=&0,\nonumber\\
\delta \rho_{\mu\nu}&=&0.
\end{eqnarray}
We can transform, with the addition of  anti-fields with opposite parity to the original fields, a set of new
auxiliary fields  $b_{\mu\nu}, B_{\mu\nu}, \overline{b}_{\mu\nu}, \overline{B}_{\mu\nu}$ which, under BRST transformations, are:
\begin{eqnarray}
\delta  g  ^{*}_{\mu\nu}&=&-b_{\mu\nu},\nonumber\\
\delta c^{*}_{\mu\nu}&=&-B_{\mu\nu},\nonumber\\
\delta \overline c^{*}_{\mu\nu}&=&-\overline{B}_{\mu\nu}, \nonumber \\
\delta L^{*}_{\mu\nu}&=&-\overline{b}_{\mu\nu}.
\end{eqnarray}
The BRST transformations of these new auxiliary fields also vanish:
\begin{eqnarray}
\delta b_{\mu\nu}&=0,\nonumber \\
\delta B_{\mu\nu}&=0,\nonumber \\
\delta \overline B_{\mu\nu}&=0,\nonumber \\
\delta \overline{b}_{\mu\nu}&=0.\label{3b}
\end{eqnarray}
Our task now is to make the tilde fields vanish by choosing a Lagrangian density to gauge fix a shift symmetry and thus recovering our original theory:
\begin{eqnarray}
\tilde {\mathcal{L}}&=& -b^{\mu\nu}\tilde{ g  }_{\mu\nu}- g  ^{*\mu \nu}(\psi_{\mu\nu}-(c_{\mu\nu}- \tilde{c}_{\mu\nu}))
-\overline{B}^{\mu\nu} \tilde{c}_{\mu\nu} + \overline{c}^{*\mu \nu}\epsilon_{\mu\nu}\nonumber \\ &&
+B^{\mu\nu}\tilde{\overline{c}}_{\mu\nu}-c^{*\mu \nu}(\overline{\epsilon}_{\mu\nu}+(L_{\mu\nu}-\tilde{L}_{\mu\nu}))+\overline{b}^{\mu\nu}\tilde{L}_{\mu\nu}+L^{*\mu \nu}\rho_{\mu\nu}.
\end{eqnarray}
By integrating out the auxiliary fields $b_{\mu\nu}, B_{\mu\nu}, \overline{B}_{\mu\nu}$ and $\overline{b}_{\mu\nu}$ the tilde fields vanish.  The resultant
 Lagrangian density is therefore invariant under the original BRST transformation and the shift transformations.  Thus, this Lagrangian density and the original Lagrangian density are both functions of the original fields.
 So we can define  $\Psi=-\overline{c}^{\mu\nu}(\mathcal{O} g  _{\mu\nu}-L_{\mu\nu}/2)$, and then, by using Eq. $(\ref{7aab})$, we can write the original Lagrangian density as
\begin{equation}
\mathcal{L}=\delta\Psi.  \label{8a}
\end{equation}
Expanding we obtain
\begin{eqnarray}
\mathcal{L}&=& \delta g  _{\mu\nu} \frac{\delta \Psi}{\delta g  _{\mu\nu}}+\delta c_{\mu\nu} \frac{\delta \Psi}{\delta c_{\mu\nu}}+\delta \overline{c}_{\mu\nu}
\frac{\delta \Psi}{\delta \overline{c}_{\mu\nu}}+\delta L_{\mu\nu}\frac{\delta \Psi}{\delta L_{\mu\nu}}\nonumber \\
&=&- \frac{\delta \Psi}{\delta g  _{\mu\nu}}\psi_{\mu\nu}+\frac{\delta \Psi}{\delta c_{\mu\nu}}\epsilon_{\mu\nu} +\frac{\delta \Psi}{\delta \overline{c}_{\mu\nu}}
\overline{\epsilon}_{\mu\nu}-\frac{\delta \Psi}{\delta L_{\mu\nu}}\rho_{\mu\nu}. \label{4b}
\end{eqnarray}
Integrating out the fields and setting the tilde to zero, we have
\begin{eqnarray}
\mathcal{L}_{\rm{tot}}&=&\tilde {\mathcal{L}}+\mathcal{L}\nonumber\\
&=&  g  ^{*\mu \nu} c_{\mu\nu}-c^{* \mu \nu}L_{\mu\nu}-\left( g  ^*  _{\mu \nu} +\frac{\delta \Psi}{\delta L^{\mu\nu}}\right)\psi^{\mu\nu} \nonumber \\
&&+\left(\overline{c}^{*}_{\mu\nu} + \frac{\delta \Psi}{\delta c^{\mu\nu}}\right)\epsilon^{\mu\nu}-\left(c^{*\mu \nu} - \frac{\delta \Psi}{\delta \overline{c}^{\mu\nu}}\right)
\overline{\epsilon}^{\mu\nu}\nonumber \\ && +\left(L^{*}_{\mu\nu} -
 \frac{\delta \Psi}{\delta L^{\mu\nu}}\right)\rho^{\mu\nu}.
\end{eqnarray}
The explicit expression for the anti-fields is obtained again in a similar manner by integrating out the ghosts associated with the shift symmetry:
\begin{eqnarray}
 g    _{\mu \nu}^*&=&-\frac{\delta\Psi}{\delta g  ^{\mu\nu}},\nonumber\\
\overline{c}  _{\mu \nu}^*&=&-\frac{\delta \Psi}{\delta c^{\mu\nu}},\nonumber\\
c^{*}_{\mu\nu}&=&\frac{\delta \Psi}{\delta \overline{c}^{\mu\nu}},\nonumber\\
L^{*}_{\mu\nu}&=&\frac{\delta \Psi}{\delta L^{\mu\nu}}.
\end{eqnarray}
This gives unique anti-fields with
\begin{eqnarray}
 g    _{\mu \nu}^*&=& \mathcal{O}\overline{c}_{\mu\nu},\nonumber\\
\overline{c}  _{\mu \nu}^*&=&0,\nonumber\\
c  _{\mu \nu}^*&=&-\mathcal{O} g  _{\mu\nu}+ \frac{L_{\mu\nu}}{2},\nonumber\\
L  _{\mu \nu}^*&=&\frac{\overline{c}_{\mu\nu}}{2}. \label{7b}
\end{eqnarray}
With these unique values we obtain an explicit form for the Lagrangian density which is invariant under the extended BRST transformations.
\section{Extended BRST  Superspace }
We now consider a superspace formalism of the previous results by using one
anti-commuting parameter $\theta$ and defining  the following superfields
\begin{eqnarray}
\varphi_{\mu\nu}(x,\theta)&=& g  _{\mu\nu}+\theta \psi_{\mu\nu},\nonumber\\
\tilde{\varphi}_{\mu\nu}(x,\theta)&=&\tilde{ g  }_{\mu\nu}+\theta(\psi_{\mu\nu}-(c_{\mu\nu}-\tilde{c}_{\mu\nu})),\nonumber\\
\chi_{\mu\nu}(x,\theta)&=&c_{\mu\nu}+\theta \epsilon_{\mu\nu},\nonumber \\
\tilde{\chi}_{\mu\nu}(x,\theta)&=&\tilde{c}_{\mu\nu}+\theta\epsilon_{\mu\nu},\nonumber \\
\overline{\chi}_{\mu\nu}(x,\theta)&=&\overline{c}_{\mu\nu}+\theta\overline{\epsilon}_{\mu\nu},\nonumber\\
\tilde{\overline{\chi}}_{\mu\nu}(x,\theta)&=&\tilde{\overline{c}}_{\mu\nu}+\theta(\overline{\epsilon}_{\mu\nu}+(L_{\mu\nu}-\tilde{L}_{\mu\nu})), \nonumber\\
f_{\mu\nu}(x,\theta)&=& L_{\mu\nu}+\theta \rho_{\mu\nu},\nonumber\\
\tilde{f}_{\mu\nu}(x,\theta)&=&\tilde{L}_{\mu\nu}+\theta \rho_{\mu\nu}.
\label{8ba}
\end{eqnarray}
Defining the following anti-superfields
\begin{eqnarray}
\tilde{\varphi}^{*}_{\mu\nu}(x,\theta)&=& g  ^{*}_{\mu\nu}-\theta b_{\mu\nu},\nonumber\\
\tilde{\chi}^{*}_{\mu\nu}(x,\theta)&=&c^{*}_{\mu\nu}-\theta B_{\mu\nu}, \nonumber \\
\tilde{\overline{\chi}}^{*}_{\mu\nu}(x,\theta)&=&\tilde{c}^{*}_{\mu\nu}-\theta \overline{B}_{\mu\nu},\nonumber \\
\tilde{f}^{*}_{\mu\nu}(x,\theta)&=& L^{*}_{\mu\nu}-\theta \overline{b}_{\mu\nu}.
\end{eqnarray}
Thus, from these superfields and anti-superfields,  we get
\begin{eqnarray}
\frac{\partial}{\partial\theta}  \tilde{\varphi^{*\mu\nu}}\tilde{\varphi}_{\mu\nu}&=&-b^{\mu\nu}\tilde{ g  }_{\mu\nu}- g  ^{*\mu \nu}(\psi_{\mu\nu}-(c_{\mu\nu}-\tilde{c}_{\mu\nu})),\nonumber\\
\frac{\partial}{\partial \theta}  \tilde{\overline{\chi}^{*\mu\nu}}\tilde{\chi}_{\mu\nu}&=&- \overline{B}^{\mu\nu}\tilde{c}_{\mu\nu}+\overline{c}^{*\mu \nu}\epsilon_{\mu\nu},\nonumber \\
-\frac{\partial}{\partial \theta}  \tilde{\overline{\chi}}^{\mu\nu}\tilde{\chi}^{*}_{\mu\nu}&=& B^{\mu\nu}\tilde{\overline{c}}_{\mu\nu}-c^{*\mu \nu}(\overline{\epsilon}_{\mu\nu}+(L_{\mu\nu}-\tilde{L}_{\mu\nu})),\nonumber\\
-\frac{\partial}{\partial \theta}\tilde{f}^{*\mu\nu}\tilde{f}_{\mu\nu}&=& \overline{b}^{\mu\nu} \tilde{L}_{\mu\nu}+L^{*\mu \nu}\rho_{\mu\nu}.
\end{eqnarray}
We can thus write in the superspace formalism the Lagrangian density given by Eq. $(\ref{3b})$ as
\begin{equation}
\tilde{\mathcal{L}}=\frac{\partial}{\partial {\theta}}\left( \tilde{\varphi}^{*\mu\nu}\tilde{\varphi}_{\mu\nu}+
\tilde{\overline{\chi}}^{*\mu\nu}\tilde{\chi}_{\mu\nu}-\tilde{\overline{\chi}}^{\mu\nu}\tilde{\chi}^{*}_{\mu\nu}-\tilde{f}^{*\mu \nu}\tilde{f}_{\mu\nu}\right ). \label{1c}
\end{equation}
Being the $\theta$ component of a superfield, this is manifestly invariant under the extended BRST transformation.  If we also consider the
gauge fixing Lagrangian density  for the original symmetry, this can also be written in this particular formalism by defining $\Phi$ as
\begin{equation}
\Phi = \Psi + \theta \delta \Psi.
\end{equation}
We therefore have
\begin{equation}
\Phi=\Psi+\theta\left(-\frac{\delta \Psi}{\delta g  ^{\mu\nu}}\psi^{\mu\nu}+\frac{\delta \Psi}{\delta c^{\mu\nu}}\epsilon^{\mu\nu}
+ \frac{\delta \Psi}{\delta \overline{c}^{\mu\nu}}\overline{\epsilon^{\mu\nu}}-\frac{\delta\Psi}{\delta L^{\mu\nu}}\rho^{\mu\nu}\right). \label{2c}
\end{equation}
Thus the original gauge-fixing Lagrangian density in the superspace formalism is given by
\begin{equation}
\mathcal{L}=\frac{\partial \Phi}{\partial \theta}. \label{4c}
\end{equation}
Thus the $\theta$ component of a superfield is again manifestly invariant under the extended BRST transformation.  The complete Lagrangian density
 can now be written as
\begin{eqnarray}
\tilde{\mathcal{L}}_{\rm{tot}}&=&\tilde{\mathcal{L}}+\mathcal{L}\nonumber \\
&=&\frac{\partial}{\partial\theta} \left (\tilde{\varphi}^{*\mu\nu}\tilde{\varphi}_{\mu\nu}+\tilde{\overline{\chi}}^{*\mu\nu}\tilde{\chi}_{\mu\nu}-
\tilde{\overline{\chi}}^{\mu\nu}\tilde{\chi}^{*}_{\mu\nu}-
\tilde{f}^{*\mu\nu}\tilde{f}_{\mu\nu}\right )+\frac{\partial{\Phi}}{\partial \theta}.
\end{eqnarray}
This Lagrangian density is manifestly invariant under the BRST symmetry, after elimination of the auxiliary and ghost fields associated with the shift symmetry.
\section{Extended Anti-BRST Lagrangian Density}
In the previous sections we analysed the extended BRST symmetry for the Lagrangian density of higher derivative a gravity theory with suitably chosen
ghost terms.  The natural extension is to discuss the extended anti-BRST symmetry of this theory.  We therefore look at the original and shifted fields which obey the extended anti-BRST transformations,
\begin{eqnarray}
\overline{\delta}\tilde{ g  }_{\mu\nu}= g  ^{*}_{\mu\nu}, && \overline{\delta} g  _{\mu\nu}=  g  ^{*}_{\mu\nu}+(c_{\mu\nu}-\tilde{\overline{c}}_{\mu\nu}), \nonumber\\
\overline{\delta}\tilde{c}_{\mu\nu}=\overline{c}^{*}_{\mu\nu}, && \overline{\delta}c_{\mu\nu}= c^{*}_{\mu\nu}+(L_{\mu\nu}-\overline{L}_{\mu\nu}), \nonumber\\
\overline{\delta}\tilde{\overline{c}}_{\mu\nu}=\overline{c}^{*}_{\mu\nu}, && \overline{\delta}\overline{c}_{\mu\nu}=\overline{c}^{*}_{\mu\nu},\nonumber\\
\overline{\delta}\tilde{L}_{\mu\nu}=L^{*}_{\mu\nu}, && \overline{\delta}L_{\mu\nu}= L^{*}_{\mu\nu}.
\end{eqnarray}
The ghost fields associated with the shift symmetry have the following extended anti-BRST transformations,
\begin{eqnarray}
\overline{\delta}\psi_{\mu\nu}&=&b_{\mu\nu}+(L_{\mu\nu}-\tilde{L}_{\mu\nu}),\nonumber\\
\overline{\delta} \epsilon_{\mu\nu}&=&B_{\mu\nu},\nonumber\\
\overline{\delta}\overline{\epsilon}_{\mu\nu}&=&\overline{B}_{\mu\nu},\nonumber\\
\overline{\delta} \rho_{\mu\nu}&=&\overline{b}_{\mu\nu}.
\end{eqnarray}
and the extended anti-BRST transformations of the anti-fields of the auxiliary fields associated with the shift symmetry vanish,
\begin{eqnarray}
\overline{\delta }b_{\mu\nu} =0, && \overline{\delta}  g  ^{*}_{\mu\nu} =0,\nonumber\\
\overline{\delta} B_{\mu\nu}=0, &&\overline{\delta} c^{*}_{\mu\nu}=0,\nonumber\\
\overline{\delta} \overline{B}_{\mu\nu}=0, &&\overline{\delta} \overline c^{*}_{\mu\nu}=0,\nonumber\\
\overline{\delta }\overline{b}_{\mu\nu}=0, && \overline{\delta} L^{*}_{\mu\nu}=0.
\end{eqnarray}
For the Lagrangian density,which is both BRST and anti-BRST invariant, it follows that it  must also be invariant under the extended anti-BRST transformation
 at least on-shell, where the transformations reduce to anti-BRST transformations.
\section{Extended Anti-BRST Superspace }
In this final section we will derive an extended BRST and an extended  on-shell anti-BRST  invariant Lagrangian density in superspace formalism. We start by
defining superfields with  two anti-commuting
parameters, namely $\theta$ and $\overline{\theta}$, as:
\begin{eqnarray}
\varphi_{\mu\nu}(x,\theta,\overline{\theta})&=& g  _{\mu\nu}+\theta \psi_{\mu\nu}+\overline{\theta}( g  ^{*}_{\mu\nu}+(\overline{c}_{\mu\nu}-\tilde{\overline{c}}_{\mu\nu}))+\theta \overline{\theta}(b_{\mu\nu}+ (L_{\mu\nu}-\tilde{L}_{\mu\nu})),
\nonumber\\
\tilde{\varphi}_{\mu\nu}(x,\theta,\overline{\theta})&=&\tilde{ g  }_{\mu\nu}+\theta(\psi_{\mu\nu}-(c_{\mu\nu}-\tilde{c}_{\mu\nu}))+\overline{\theta} g  ^{*}_{\mu\nu}+\theta \overline{\theta}b_{\mu\nu},
\nonumber\\
\chi_{\mu\nu}(x,\theta,\overline{\theta})&=&c_{\mu\nu}+\theta \epsilon_{\mu\nu}
+\overline{\theta}(c^{*}_{\mu\nu}+(L_{\mu\nu}-\tilde{L}_{\mu\nu}))+\theta\overline{\theta}B_{\mu\nu},
 \nonumber\\
\tilde{\chi}_{\mu\nu}(x,\theta,\overline{\theta})&=&\tilde{c}_{\mu\nu}+\theta \epsilon_{\mu\nu}+\overline{\theta}c^{*}_{\mu\nu}+\theta\overline{\theta}B_{\mu\nu},
\nonumber\\
\overline{\chi}_{\mu\nu}(x,\theta,\overline{\theta})&=&\overline{c}_{\mu\nu}+\theta\overline{\epsilon}_{\mu\nu}+\overline{\theta} \overline{c}^{*}_{\mu\nu} +\theta \overline{\theta}\overline{B}_{\mu\nu},
\nonumber\\
\tilde{\overline{\chi}}_{\mu\nu}(x,\theta,\overline{\theta})&=&\tilde{\overline{c}}_{\mu\nu}+\theta(\overline{\epsilon}_{\mu\nu}+(L_{\mu\nu}-\tilde{L}_{\mu\nu})) + \overline{\theta}\overline{c}^{*}_{\mu\nu}
+\theta \overline{\theta}\overline{B}_{\mu\nu}. \label{7c}
\end{eqnarray}
We therefore have
\begin{eqnarray}
 -\frac{1}{2}\frac{\partial}{\partial \overline{\theta}}\frac{\partial}{\partial \theta} \tilde{\varphi}^{\mu\nu}\tilde{\varphi}_{\mu\nu}&=&- b^{\mu\nu}\tilde{ g  }_{\mu\nu}
- g  ^{*\mu}(\psi_{\mu\nu}-(c_{\mu\nu}-\tilde{c}_{\mu\nu})),\\
\frac{\partial}{\partial \overline{\theta}}\frac{\partial}{\partial \theta}\tilde{\chi}^{\mu\nu}\tilde{\overline{\chi}}_{\mu\nu}&=&
- \overline{B}^{\mu\nu}\tilde{c}_{\mu\nu}
+\overline{c}^{*\mu\nu}\epsilon_{\mu\nu} +B^{\mu\nu}\tilde{\overline{c}}_{\mu\nu} \nonumber \\ && -c^{*\mu\nu}
(\overline{\epsilon}_{\mu\nu}+(L_{\mu\nu}-\tilde{L}_{\mu\nu})).
\end{eqnarray}
Therefore
\begin{eqnarray}
 \tilde{\mathcal{L}}&=&\frac{\partial}{\partial \overline{\theta}}\frac{\partial}{\partial \theta}
-\frac{1}{2}\tilde{\varphi}^{\mu\nu}\tilde{\varphi}_{\mu\nu}+\tilde{\chi}^{\mu\nu}\tilde{\overline{\chi}}_{\mu\nu}\nonumber \\ &=&
- b^{\mu\nu}\tilde{ g  }_{\mu\nu}- g  ^{*\mu\nu}(\psi_{\mu\nu}-(c_{\mu\nu}-\tilde{c}_{\mu\nu}))-\overline{B}^{\mu\nu}\tilde{c}_{\mu\nu}+\overline{c}^{*\mu}\epsilon_{\mu\nu} \\ && +B^{\mu\nu}\tilde{\overline{c}}_{\mu\nu}-c^{*\mu\nu}(\overline{\epsilon}_{\mu\nu}+(L_{\mu\nu}-\tilde{L}_{\mu\nu})).
\end{eqnarray}
Being the $\theta \overline{\theta}$ component of a superfield, this gauge-fixing Lagrangian density is manifestly invariant under extended BRST and anti-BRST
 transformations. Furthermore, we define
\begin{equation}
\Phi(x,\theta,\overline{\theta})=\Psi+\theta \delta \Psi+ \overline{\theta}\overline{\delta}\Psi
+\theta \overline{\theta} \delta \overline{\delta}\Psi. \label{3d}
\end{equation}
The component of $\theta \overline{\theta}$ can be made to vanish on-shell and therefore the Lagrangian density for the original
 fields can be written as
\begin{equation}
\mathcal{L}=\frac{\partial}{\partial \theta}(\delta(\overline{\theta}) \Phi(x, \theta, \overline{\theta})). \label{7d}
\end{equation}
This Lagrangian density is both manifestly invariant under extended BRST transformations and invariant under extended anti-BRST transformations
 on-shell.  The complete Lagrangian density is therefore
\begin{eqnarray}
\mathcal{L}_{\rm{tot}}&=&\tilde{\mathcal{L}} +\mathcal{L}\nonumber\\
&=&\frac{\partial}{\partial \overline{\theta}} \frac{\partial}{\partial \theta}
\left(-\frac{1}{2}\tilde{\varphi}^{\mu\nu}\tilde{\varphi}_{\mu\nu}+
 \tilde{\chi}^{\mu\nu}\tilde{\overline{\chi}}_{\mu\nu} \right) +
\frac{\partial}{\partial \theta}(\delta(\overline{\theta})\Phi(x,\theta,\overline{\theta}))\nonumber\\
&=&- b^{\mu\nu}\tilde{ g  }_{\mu\nu}-\overline{B}^{\mu\nu}\tilde{c}_{\mu\nu}+B^{\mu\nu}\tilde{\overline{c}}_{\mu\nu} 
- \left( g  ^{*}_{\mu\nu}+\frac{\delta \Psi}{\delta  g
^{\mu\nu}}\right)\psi^{\mu\nu} \nonumber \\ && + g  ^{*\mu\nu}(c_{\mu\nu}-\tilde{c}_{\mu\nu})-c^{*\mu\nu}(L_{\mu\nu}-\tilde{L}_{\mu\nu})
 \nonumber\\
&& +\left(\overline{c}^{*\mu\nu}
 + \frac{\delta \Psi_{\mu\nu}}{\delta c^{\mu\nu}}\right)\epsilon^{\mu\nu}-\left(c^{*\mu\nu}-
\frac{\delta \Psi}{\delta \overline{c}^{\mu\nu}}\right)\overline{\epsilon}^{\mu\nu}. \label{8d}
\end{eqnarray}
The tilde fields vanish when we integrate out the auxiliary fields.  Additionally
by integrating out the ghost fields for the shift symmetry we will get explicit expressions for the antifields.
As $L_{\mu\nu}$ and $\tilde{L}_{\mu\nu}$ are auxiliary fields, so we can redefine them as $L_{\mu\nu}-\tilde{L}_{\mu\nu}\to L_{\mu\nu}$.
This combination $(L_{\mu\nu}+\tilde{L}_{\mu\nu})$ can now  be integrated out and absorbed into the normalization constant.
Thus we have obtained a  Lagrangian density in superspace
formalism which is manifestly BRST invariant and also manifestly anti-BRST invariant on-shell.
\section{Conclusion}
We have studied the anti-BRST and double BRST symmetries of  perturbative higher derivative gravity. We have also analysed the extended BRST and the extended anti-BRST
symmetries of this theory in the superspace formalism.
 This theory  was found to be invariant under extended BRST transformations. However on extended anti-BRST  transformations it was found to be only invariant on-shell.

 One may develop a supersymmetric version of this theory and thus  study   certain super-gravity theories with higher
derivative terms.  The BRST symmetry associated with the higher derivative structure of such theories has not been studied.
We can proceed to construct
such a theory by considering the theory studied here to be the bosonic part of that  higher derivative super-gravity theory. 
 We could generalise the results of this paper to curved spacetime. In particular one could analyse perturbative higher derivative
 gravity in de Sitter and anti-de Sitter spacetimes \cite{desitter}-\cite{desitter1}.
It will also be interesting to investigate higher derivatives in gravity using
 Wheeler-DeWitt equation
 \cite{zp1}-\cite{zq}.

\end{document}